# Performance of Conventional EEG Biomarkers Across Different Clinical Phases of Major Depressive Disorder: A Comprehensive Evaluation

Feng Yan[#,1,2], Xuteng Wang[#,1,2], Shuyi Yang[3], Yue Zhao[1,2], Xiaobin Wong[#,4,+], Zhiren Wang[1,2,+]

1: Beijing HuiLongGuan Hospital, Capital Medical University

2: Peking University HuiLongGuan Clinical Medical School

3: Xinya College, Tsinghua University

4: AI Innovation Center, Tsinghua Pearl River Delta Research Institute

#: equal contribution

+: corresponding author

## Abstract

While EEG features differentiate Major Depressive Disorder (MDD) from healthy controls (HC), their clinical utility as biomarkers depends on a monotonic trajectory across the disease spectrum—from the acute (AC) phase to the maintenance (MA) phase and finally to the healthy baseline. However, the progression of the MA phase remains poorly understood in traditional marker analysis. Analyzing EEG data from 74 individuals (24 AC, 23 MA, and 27 HC), this study provides a comprehensive evaluation of classic ERP and resting-state indices across AC, MA, and HC groups. Our results demonstrate that almost no conventional metrics strictly satisfy the criterion of monotonic progression, likely due to profound inter-individual heterogeneity. These findings highlight the inherent limitations of group-level feature extraction and provide critical insights for developing future paradigms and algorithms to identify neurobiological markers with genuine clinical utility.

## 1. Introduction

Major depressive disorder (MDD) has become one of the most serious global public health challenges of the 21st century, affecting approximately 3.8% of the world's population (Organization, 2017). This disorder not only severely impairs patients' quality of life and social functioning but is also closely associated with significant cognitive deficits (Malhi & Mann). Currently, the clinical diagnosis of depression primarily relies on symptom-based assessments and psychometric scales. However, these approaches suffer from several limitations, including high subjectivity and clinician experience (Marx et al., 2023). Therefore, developing objective biomarkers grounded in neurobiological mechanisms holds substantial clinical value for the early detection and precise diagnosis.

Abnormalities in emotional processing are a core feature of MDD, manifested as attentional bias toward negative emotional stimuli, diminished experience of positive emotions, and impaired emotion regulation (Beck & Haigh, 2014). Using electroencephalography (EEG) makes it particularly suitable for investigating rapid emotional and cognitive processes (Costa et al., 2014). Currently, event-related potentials (ERPs) has provided rich evidence of altered emotional and cognitive processing in patients with MDD. In emotional word or picture processing paradigms, individuals with MDD often exhibit enhanced early negative components (such as P1 or N1) in response to negative stimuli, reflecting heightened automatic attention to negative information (Bruder et al., 2011; Foti & Hajcak, 2009). Abnormalities have also been reported in P2 and N2 components, which are associated with early affective discrimination and conflict monitoring. Patients with MDD frequently show reduced P2 amplitudes to positive stimuli and enhanced N2 responses to negative stimuli, indicating biased emotional appraisal and impaired inhibitory control (Li et al., 2018; Liu et al., 2014). However, findings from existing studies remain inconsistent.(Bruun et al., 2021; Kovacevic et al., 2025)

Apart from stimulus-evoked perspective, resting-state electroencephalography (rsEEG) analyses have also demonstrated marked alterations in functional patterns between patients with MDD and healthy controls. Specifically, MDD patients often exhibit abnormal synchronization across frontal and parietal regions, reduced alpha-band coherence, and altered small-world properties of brain networks, suggesting disrupted large-scale neural communication underlying depressive symptoms (Fingelkurts et al., 2007; Greicius et al., 2007; Kaiser et al., 2015). These findings suggest that depression is characterized by a pervasive negative bias across multiple stages of emotional information processing, accompanied by a sustained aberrant neural state.

However, in addition to the inconsistencies observed across various studies, most of the studies examining the differences between patients with MDD and healthy individuals have not taken into account the changes that occur during the recovery phase of depression so called continuation phase or maintenance phase(Cavanagh et al., 2019). Ideally, a valid biomarker should be prominently expressed in patients during the acute phase, gradually diminish as patients recover, and remain absent in healthy individuals. Such biomarkers hold greater potential for translation into clinical decision support.

The present study focuses on evaluating the conventional EEG biomarker differences among patients at different stages of depression and healthy control, based on both an emotional word stimulation paradigm (task-evoked) and resting-state (task-free) analyses. In addition, to better evaluate the effectiveness of different biomarkers in assessing disease progression.

## 2. MATERIAL AND METHODS

## 2.1 Participants

There is an overlap in participants between this dataset and one of our previous studies(Yan et al., 2025). Finally, 74 participants (AC: 24，MA: 23，HC：27) was involved in this study.

Table 1: Participants in ERP analysis:

| Groups | Gender | Age (mean±SD) | BDI (mean±SD) | SAS (mean±SD) | HAMD (mean±SD) | HAMA (mean±SD) |
| --- | --- | --- | --- | --- | --- | --- |
| AC | 17F7M | 19.42±5.44 | 26.80±12.28 | 61.08±12.15 | 24.54±6.51 | 25.29±8.78 |
| MA | 17F6M | 31.1±11.37 | 7.22±6.00 | 40±9.04 | 8.91±4.20 | 9.74±6.00 |
| HC | 17F10M | 24.81±4.09 | 5.25±5.91 | 36.70±7.90 | 4.30±5.11 | 4.29±6.11 |

AC: Acute phase group;
MA: Maintenance phase group;
HC: Healthy Controls group
BDI: the Beck Depression Inventory (BDI)
SAS: the Self-Rating Anxiety Scale (SAS)
HAMD: the Hamilton Depression Rating Scale (HAMD-17)
HAMA: the Hamilton Anxiety Rating Scale (HAMA)

In this research, all patients with MDD met the DSM-5 diagnostic criteria for a depressive episode. Among them, the acute phase group (AC) was defined as the period—typically within 12 weeks following the onset of a major depressive episode—in which the primary clinical objective of treatment is symptom reduction and the restoration of social functioning, with a Hamilton Depression Rating Scale (HAMD) score exceeding 16. Given the clinical difficulty in distinguishing between the maintenance and continuation phases, this study unified these periods into a single definition: the interval between 6 and 24 months following the resolution of acute symptoms. During this period, the primary clinical goal is the prevention of a new, independent depressive episode, and the HAMD score remains less than or equal to 16. These people was marked as maintenance phase group (MA) in this study. Individuals with a history of head injury, schizophrenia, or current pregnancy were excluded from the study.

Each participant underwent a sequence of resting-state recordings: 2 minutes with eyes open, followed by 2 minutes with eyes closed. Then the task (if have) will be enrolled in. After that, repeated to obtain a full cycles (open-closed), ensuring sufficient relaxation and physiological stabilization. Participants were instructed to press a button to advance to each subsequent stage, and an auditory cue was presented to signal the end of each eyes-closed interval.

## 2.2 Task and materials

The experimental task employed an emotional word stimulus paradigm. Stimuli were derived from the AFINN sentiment lexicon(Nielsen, 2011) and underwent rigorous cross-cultural adaptation and Chinese translation.

Positive words (e.g., "happy," "bliss," "success") had valence scores ranging from +2 to +3, while negative words (e.g., "pain," "failure," "despair") were scored between -3 and -2. Neutral words (e.g., "adopt," "transparent," "resolve") fell within the range of -1 to +1. To minimize interference from socio-cognitive factors, words with prominent social attributes were excluded.

An attentional control task was implemented, requiring participants to identify specific Chinese names embedded within the emotional words to ensure sustained attention to all stimuli. A Latin square design was utilized to counterbalance conditions and mitigate potential order and fatigue effects. Each word was presented for 2 seconds, with an 0.5 seconds inter-stimulus interval (ISI).

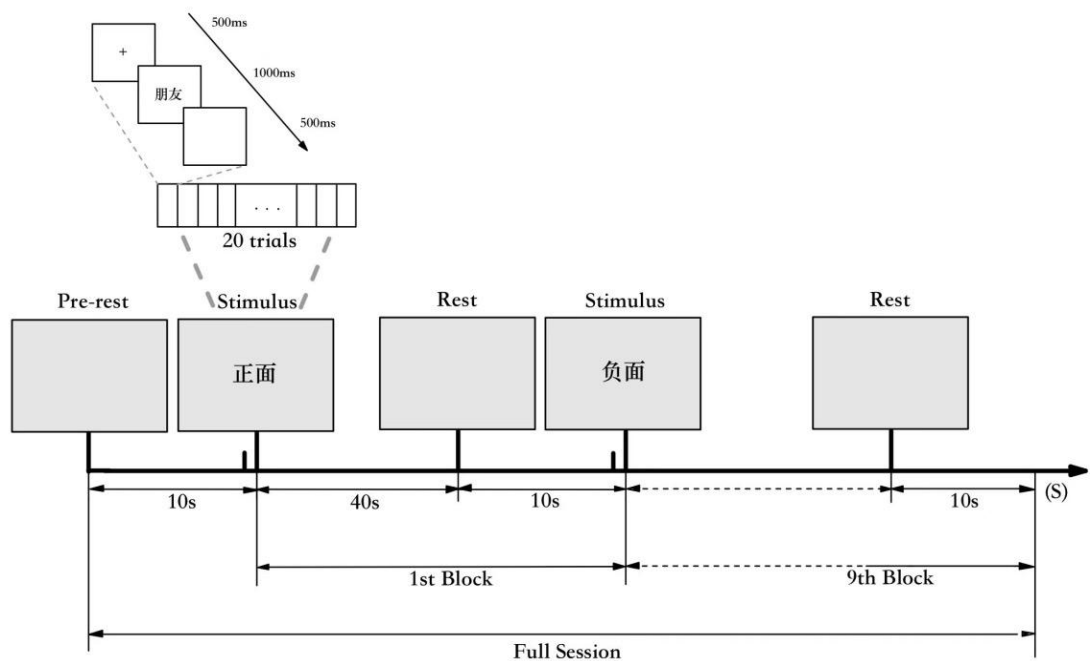

(Figure1: Task-based Experimental Procedure. A total of 180 two-character Chinese words were selected as stimuli, comprising the following categories: 60 Positive ["正面"] words: Valence scores ranging from +2 to +3, selected after professional translation (e.g., "快乐" [happiness], "幸福" [bliss], "成功" [success]). 60 Negative words["负面"] : Valence scores ranging from -3 to -2 (e.g., "痛苦" [pain], "失败" [failure], "绝望" [despair]). 60 Neutral words: Valence scores ranging from -1 to +1 (e.g., "采纳" [adopt], "透明" [transparent], "解决" [resolve]). In addition, 20 common Chinese names were selected as target stimuli for evaluating the attentional involved. A Latin square design was implemented for both within-subject and between-subject factors to ensure rigorous counterbalancing.)

The entire experiment took approximately 40 minutes while the task took 20. The study was approved by the Institutional Medicine Review Board of Beijing Huilongguan Hospital (approval number:2024-106-research).

## 3. Data Analysis

3.1 EEG Recording and preprocessing

EEG data preprocessing followed the methods described in the previous study (Yan et al., 2025). EEG signals were recorded using a wireless EEG acquisition system (NeuSen.W32, Neuracle, China) at a sampling rate of 1000 Hz. Thirty-two saline-based sponge electrodes were positioned according to the international 10–20 system. Electrode impedance was maintained below 30 kΩ throughout the experiment to ensure acceptable quality signal acquisition.

The experimental paradigm was developed using PsychoPy (version 2022.2.5). EEG data were analyzed using Python (version 3.8.20) and the MNE-Python package (version 1.6.1). The preprocessing pipeline included: removing reference channels (A1/A2), band-pass filtering (1–40 Hz), notch filtering (50 Hz), interpolation of bad channels, automated Independent Component Analysis (ICA) for ocular artifact removal, and re-referencing to the common average(Chen et al., 2023). All analysis scripts are open-source and available at:
https://disk.pku.edu.cn/link/AA10BC087C31424B6E90F81161619E9143

## 3.2 Event-related analysis

A 3 (Group: Healthy Control, Acute Phase (AC), Maintenance Phase (MA)) × 3 (Valence: Positive, Neutral, Negative) mixed factorial design was employed, resulting in nine experimental conditions. For each trial, baseline correction was performed using the 500 ms pre-stimulus interval relative to the 1500 ms post-stimulus period.

Event-Related Potential (ERP) components were extracted based on their representative regions of interest (ROIs) and standardized time windows:

**N1/P1** (Early Visual Attention): Measured over occipital electrodes ('O1', 'O2', 'PO3', 'PO4') within a 50–150 ms window.

**N170** (Orthographic Processing & Early Emotional Sensitivity): Extracted from temporo-parietal sites ('P7', 'P8') between 140–180 ms.

**EPN** (Early Posterior Negativity): Analyzed at posterior locations ('P7', 'P8', 'PO3', 'PO4') within 200–300 ms.

**P2**: Measured at fronto-central sites ('Cz', 'FC1', 'FC2') during 150–250 ms.

**N2**: Extracted from fronto-central electrodes ('FC1', 'FC2') within 200–350 ms.

**P300**: Focused on the central site ('Cz') within a 250–350 ms window.

Mean amplitudes for each component were calculated within their respective time windows for subsequent statistical comparisons.

## 3.3 Resting-state analysis

Resting-state data were categorized into eyes-open (EO) and eyes-closed (EC) conditions. Power Spectral Density (PSD) was estimated using Welch's method with 5-second non-overlapping epochs. To obtain a representative profile for each subject, PSD values were averaged across all available epochs. Statistical analyses were subsequently performed on the mean power within predefined frequency bands to evaluate group-specific oscillatory patterns.

This study also executes the brain network analysis. For the brain network analysis, connectivity matrices were binarized using a proportional thresholding (density = 20%) to retain the most robust functional connections. We computed both global and local topological metrics using the NetworkX and Pyintergraph libraries, including mean node strength, clustering coefficient, global efficiency, and modularity. To assess the optimal balance between functional integration and segregation, the small-worldness index ($\sigma$) was estimated by comparing the empirical graphs to 20 randomized surrogate networks. Statistical differences across the three groups (HC, MA, AC) were evaluated using Kruskal-Wallis tests, with post-hoc effect sizes estimated via Cohen's d and p-values corrected for multiple comparisons using bonferroni.

## 4. Results

In this study, Kruskal-Wallis tests were consistently employed to assess group differences, followed by post-hoc pairwise comparisons for conditions exhibiting significant or marginal effects.

Regarding the N170 component, a marginal group effect was observed for both neutral ($H(2) = 7.039, p_{bonf} = 0.089$) and negative ($H(2) = 7.134, p_{bonf} = 0.084$) valences. Post-hoc analysis revealed that N170 amplitudes in the AC group were significantly higher than those in the HC group under both neutral ($p_{adj} = 0.028$) and negative ($p_{adj} = 0.050$) conditions.

For the P300 component, a marginal main effect was found in the positive condition ($H(2) = 8.060, p_{bonf} = 0.053$), where the AC group exhibited significantly larger amplitudes compared to both the HC ($p_{adj} = 0.047$), $MA (p_{adj} = 0.042)$ groups.

Additionally, a significant group difference was identified for the N2 component in the positive condition ($H(2) = 8.228, p_{bonf} = 0.049$). Post-hoc tests indicated that the AC group's N2 amplitude was significantly larger than that of the HC group ($p_{adj} = 0.017$).

In the resting-state analysis, significant or marginal group differences were identified exclusively in the prefrontal regions during the eyes-closed condition. Specifically, spectral power in the delta ($H(2) = 6.428, p_{bonf} = 0.080$) and theta ($H(2) = 6.221, p_{bonf} = 0.044$) bands exhibited statistical variations. Post-hoc comparisons revealed that the power spectral density (PSD) in the AC group was generally higher than that in the MA phase (delta: $p_{adj} = 0.050$; theta: ($p_{adj} = 0.078$), whereas no significant differences were observed between the AC group and the HC group.

Graph theoretical analysis revealed significant group differences in mean node strength across all frequency bands ($p_{bonf} < 0.01$). Post-hoc comparisons indicated that the HC group exhibited significantly higher network connectivity strength compared to both the AC and MA phase patients. Notably, network strength in the MA group showed a further declining trend, with mean values across all bands consistently lower than those of the AC group. Regarding mean clustering coefficient, significant group effects were observed in all frequency bands ($p_{bonf} < 0.05$), with the exception of the delta band, which showed a marginal trend ($p_{bonf} = 0.068$). The overall pattern of results for mean clustering was highly consistent with that of the mean node strength.

## 5. Discussion

As demonstrated in the results, several EEG biomarkers reached statistical significance ($p < 0.05$) across different groups and conditions. In this section, rather than extensively exploring the underlying neurobiological mechanisms of these differences, we focus on their potential as indicators of disease progression. Specifically, we examine whether these biomarkers strictly follow the hypothesized gradient, where the values of the MA group are positioned between those of the AC and HC groups. Figure 2 presents the boxplots for all biomarkers that showed significant differences specifically between the AC and HC groups.

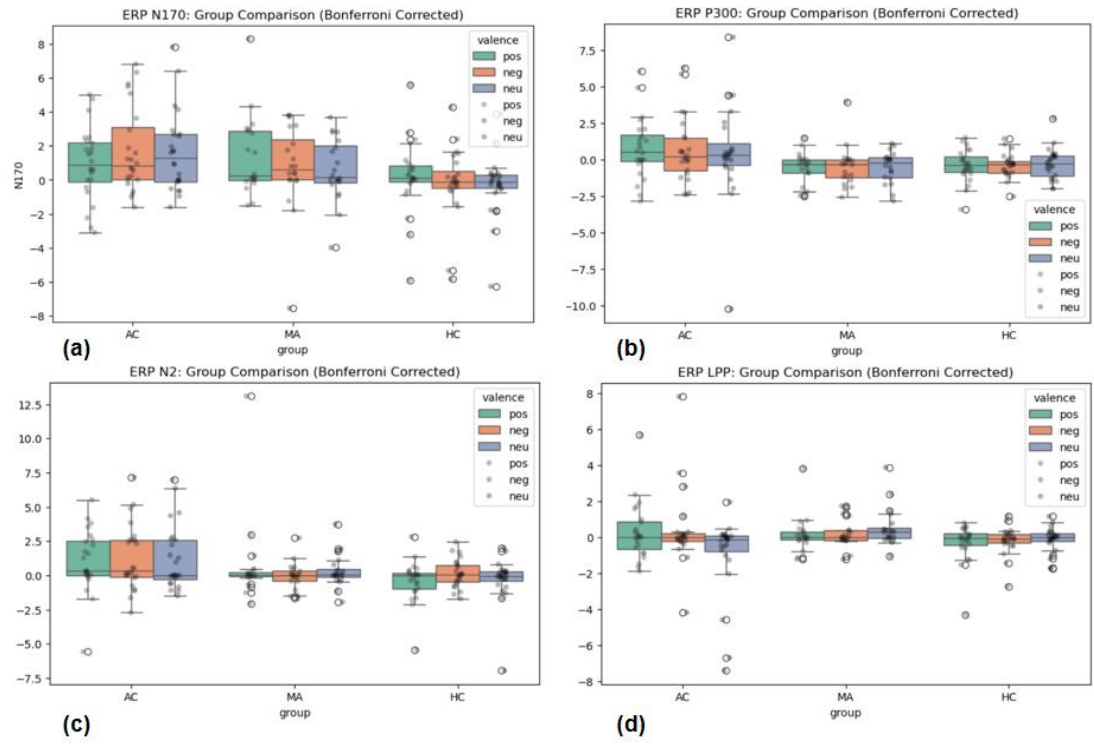

(Figure2: ERP components across different stages and conditions: (a) N170, (b) P300, (c) N2, and (d) LPP. Only components exhibiting significant group differences in the results are displayed. In the search for biomarkers to evaluate disease progression and recovery, features whose values in the maintenance (MA) group fall between those of the acute (AC) and healthy control (HC) groups are of particular interest. Such characteristics provide a robust metric for longitudinal monitoring; as a patient progresses through recovery, a shift in these values toward the healthy baseline can effectively indicate clinical improvement.)

As illustrated in Figure 2, only the N170 and P300 components, especially the mean amplitude of different emotions in N170, demonstrate the anticipated trend, with the characteristic values of the MA group positioned between those of the AC and HC groups. These findings suggest that among traditional ERP indices, N170 might serve as biomarkers with substantial potential for evaluating clinical progression and recovery trends in MDD.

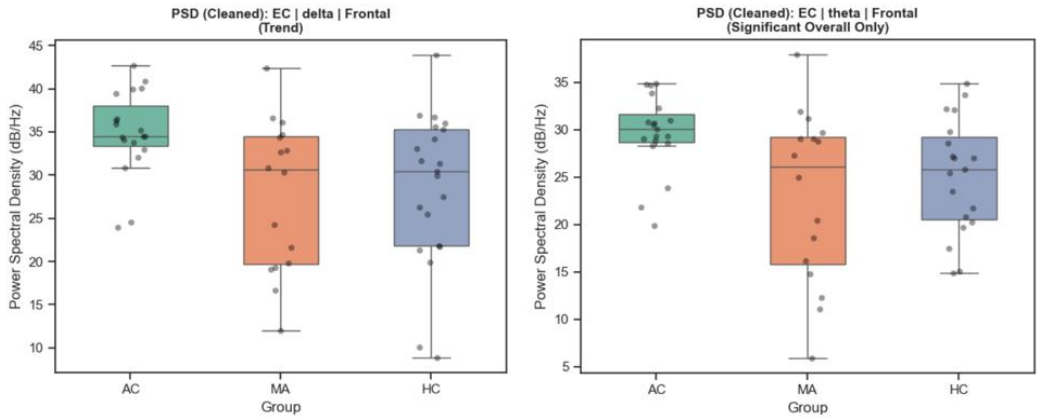

(Figure3: Boxplots showing the mean PSD features in different brain region across different frequency bands in different groups during the eyes-closed resting state.)

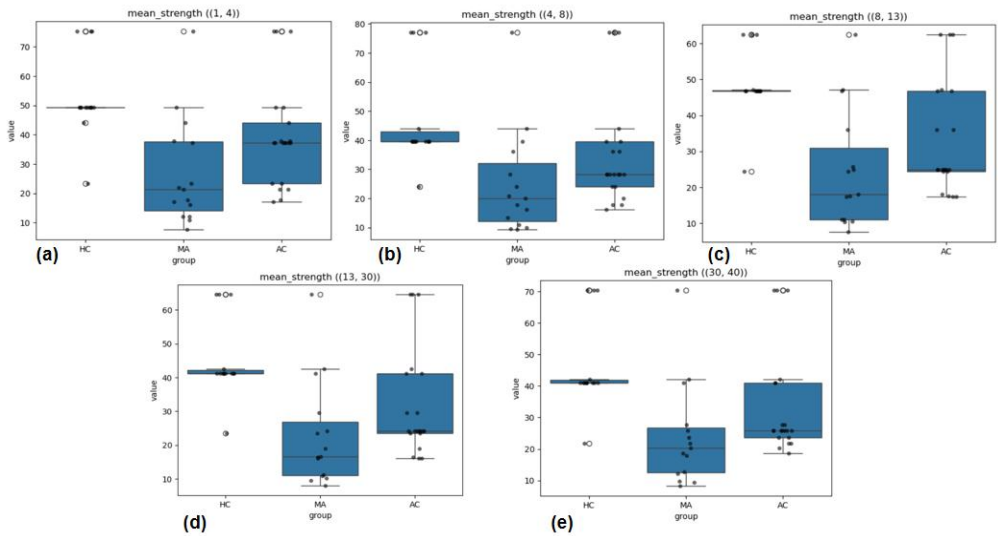

(Figure4: Boxplots showing the mean whole-brain functional connectivity strength across different frequency bands in different groups during the eyes-closed resting state. From (a) to (e), none of the frequency bands conformed to the hypothesis that the mean values for the MA group would fall between those of the AC and HC groups.)

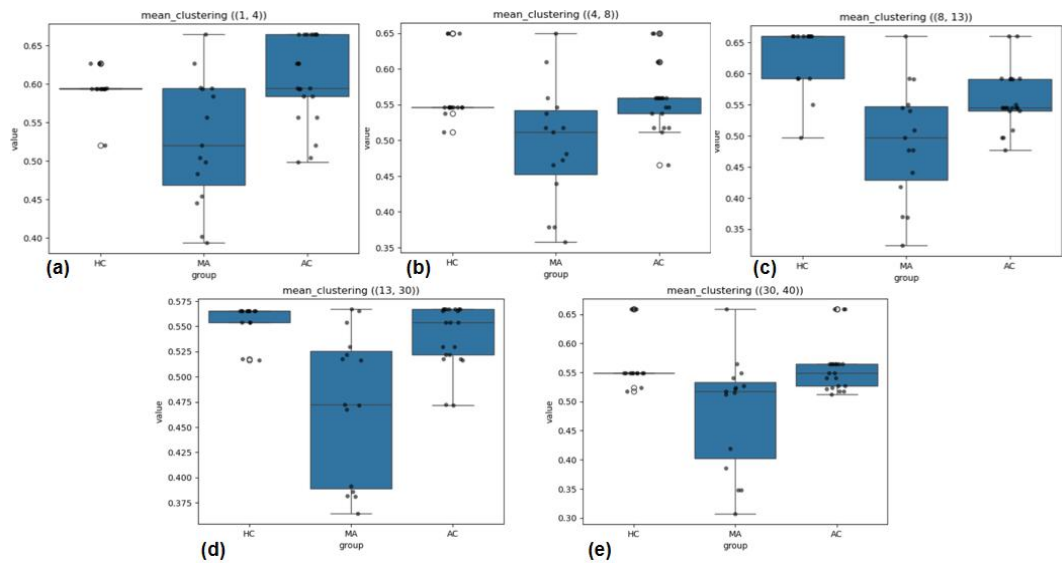

(Figure5: Boxplots showing the mean whole-brain functional connectivity clustering coefficient across different frequency bands in different groups during the eyes-closed resting state. From (a) to (e), none of the frequency bands conformed to the hypothesis that the mean values for the MA group would fall between those of the AC and HC groups.)

As illustrated in Figures 3–5, nearly all resting-state features fail to conform to the hypothesized trend of a progressive shift from AC toward HC during the recovery process. Instead, these metrics exhibit non-linear patterns characterized by abrupt increases or decreases during the maintenance phase. While complex neurobiological mechanisms may underlie these observations, such patterns pose significant challenges for clinical application. Specifically, they create diagnostic ambiguity: if a characteristic value increases, it remains unclear whether this indicates clinical improvement toward the MA stage or deterioration toward the AC stage. Consequently, these features currently lack the necessary reliability to serve as clinically viable biomarkers for monitoring disease progression.

However, it is worth noting that nearly all identified biomarkers exhibit substantial variance. This high degree of inter-subject variability suggests a fundamental limitation in seeking biomarkers solely through cross-sectional observations and group-averaging approaches. An emerging perspective in Precision Psychiatry suggests that such group-level inconsistencies are often masked by profound inter-individual and intra-individual heterogeneity. (Fernandes et al., 2017; Wong et al., 2026). Consequently, a growing number of researchers have shifted their focus toward overcoming the challenges posed by individual variability, aiming to extract more robust and reliable biomarkers from these complex neurophysiological patterns(Wong et al., 2026).

## Conclusion

This study evaluated a range of EEG features previously reported to differentiate Major Depressive Disorder (MDD) from Healthy Controls (HC), aiming to validate these potentially controversial biomarkers within our own clinical dataset to quantify disease progression.

We sought to identify biomarkers that exhibit a monotonic trajectory transitioning from the acute (AC) phase to the maintenance (MA) phase and finally to the healthy (HC) baseline, which would provide more objective clinical guidance for evaluating recovery. Our comprehensive approach integrated both task-related ERPs (via an emotional Stroop paradigm) and resting-state characteristics (including power spectral density and graph network metrics across eyes-open and eyes-closed conditions), evaluating these indices within a holistic diagnostic framework.

Our results indicate that almost no conventional metrics strictly satisfy this criterion, highlighting the inherent limitations of traditional group-level feature extraction in capturing the complexities of MDD recovery. **This discrepancy unmasks the fundamental constraints of group-averaging approaches and the inescapable challenge of inter-individual variability. Furthermore, it offers a pivotal perspective for designing benchmarks tailored to Precision Psychiatry: specifically, shifting the focus toward longitudinal intra-individual tracking and repeated assessments throughout the recovery process to identify robust biomarkers or moderators.**

**Data Availability Statement**

The preprocessed data and code supporting the conclusions of this article will be made available upon reasonable request. All the data will be shared after the whole collection project has been completed.

**Ethics Statement**

The studies involving human participants were approved by the the Institutional Medicine Review Board of Beijing Huilongguan Hospital (approval number:2024-106-research). All studies were conducted in accordance with local legislation and institutional requirements. Written informed consent was obtained from all participants.

**Code Availability**

All code used for data preprocessing and technical validation is available at: https://disk.pku.edu.cn/link/AA10BC087C31424B6E90F81161619E9143

**Acknowledgements**

This work was supported by the Capital's Funds for Health Improvement and Research (2022-2-2133, 2024-1-2131).

**Author Contributions**
Conceptualization: XBW, FY, XTW
Clinical Support: FY, XTW, YZ
Methodology: XBW, SYY, XTW
Investigation: ZRW
Visualization: XTW, SYY, YZ
Supervision: ZRW
Writing—Original Draft: XBW, FY, XTW
Writing—Review & Editing: ZRW, XBW

**Competing Interests**

Authors declare that they have no competing interests.

# Reference:

Beck, A. T., & Haigh, E. A. (2014). Advances in cognitive theory and therapy: The generic cognitive model. *Annual review of clinical psychology*, *10*(1), 1–24.

Bruder, G. E., Kayser, J., & Tenke, C. E. (2011). Event-related brain potentials in depression: Clinical, cognitive, and neurophysiological implications.

Bruun, C. F., Arnbjerg, C. J., & Kessing, L. V. (2021). Electroencephalographic parameters differentiating melancholic depression, non-melancholic depression, and healthy controls. a systematic review. *Frontiers in psychiatry*, *12*, 648713.

Cavanagh, J. F., Bismark, A. W., Frank, M. J., & Allen, J. J. B. (2019). Multiple Dissociations Between Comorbid Depression and Anxiety on Reward and Punishment Processing: Evidence From Computationally Informed EEG. *Comput Psychiatr*, *3*, 1–17. https://doi.org/10.1162/cpsy_a_00024

Chen, J., Wang, X., Huang, C., Hu, X., Shen, X., & Zhang, D. (2023). A Large Finer-grained Affective Computing EEG Dataset. *Sci Data*, *10*(1), 740. https://doi.org/10.1038/s41597-023-02650-w

Costa, T., Cauda, F., Crini, M., Tatu, M.-K., Celeghin, A., de Gelder, B., & Tamietto, M. (2014). Temporal and spatial neural dynamics in the perception of basic emotions from complex scenes. *Social cognitive and affective neuroscience*, *9*(11), 1690–1703.

Fernandes, B. S., Williams, L. M., Steiner, J., Leboyer, M., Carvalho, A. F., & Berk, M. (2017). The new field of 'precision psychiatry'. *BMC Med*, *15*(1), 80. https://doi.org/10.1186/s12916-017-0849-x

Fingelkurts, A. A., Fingelkurts, A. A., Rytsälä, H., Suominen, K., Isometsä, E., & Kähkönen, S. (2007). Impaired functional connectivity at EEG alpha and theta frequency bands in major depression. *Human brain mapping*, *28*(3), 247–261.

Foti, D., & Hajcak, G. (2009). Depression and reduced sensitivity to non-rewards versus rewards: Evidence from event-related potentials. *Biological psychology*, *81*(1), 1–8.

Greicius, M. D., Flores, B. H., Menon, V., Glover, G. H., Solvason, H. B., Kenna, H., Reiss, A. L., & Schatzberg, A. F. (2007). Resting-state functional connectivity in major depression: abnormally increased contributions from subgenual cingulate cortex and thalamus. *Biological psychiatry*, *62*(5), 429–437.

Kaiser, R. H., Andrews-Hanna, J. R., Wager, T. D., & Pizzagalli, D. A. (2015). Large-scale network dysfunction in major depressive disorder: a meta-analysis of resting-state functional connectivity. *JAMA psychiatry*, *72*(6), 603–611.

Kovacevic, N., Meghdadi, A., Berka, C., Saad, Z. S., Kolb, H. C., de Boer, P., Furey, M., & Miller, S. (2025). Differences in resting state and task-based EEG measures between patients with major depressive disorder and healthy controls. *Clinical Neurophysiology*, *173*, 190–198.

Li, X., Li, J., Hu, B., Zhu, J., Zhang, X., Wei, L., Zhong, N., Li, M., Ding, Z., & Yang, J. (2018). Attentional bias in MDD: ERP components analysis and classification using a dot-probe task. *Computer methods and programs in biomedicine*, *164*, 169–179.

Liu, H., Yin, H.-f., Wu, D.-x., & Xu, S.-j. (2014). Event-related potentials in response to emotional words in patients with major depressive disorder and healthy controls. *Neuropsychobiology*, *70*(1), 36–43.

Malhi, G. S., & Mann, J. J. Depression Lancet, 392 (10161)(2018). *View PDF View article View in Scopus*, 2299–2312.

Marx, W., Penninx, B. W., Solmi, M., Furukawa, T. A., Firth, J., Carvalho, A. F., & Berk, M. (2023). Major depressive disorder. *Nature Reviews Disease Primers*, *9*(1), 44.

Nielsen, F. Å. (2011). A new ANEW: Evaluation of a word list for sentiment analysis in microblogs. *arXiv preprint arXiv:1103.2903*.

Organization, W. H. (2017). Depression and other common mental disorders: global health estimates. In *Depression and other common mental disorders: global health estimates*.

Wong, X., Zhao, Z., Guo, H., Liu, Z., Wu, Y., Yan, F., Wang, Z., & Song, S. (2026). CRCC: Contrast-Based Robust Cross-Subject and Cross-Site Representation Learning for EEG. *arXiv preprint arXiv:2602.19138*.

Yan, F., Wang, X., Zhao, Y., Yang, S., & Wang, Z. (2025). Differences in Neurovascular Coupling in Patients with Major Depressive Disorder: Evidence from Simultaneous Resting-State EEG-fNIRS. *arXiv preprint arXiv:2506.11634*.